\documentclass{aa}
\usepackage{graphicx}
\usepackage{natbib}
\begin{document}

\def\ctq{CTQ~414}
\def\bdix{B~1030+074}
\def\hezero{HE~0230-2130}
\def\heonze{HE~1104-1805}
\def\lbqs{LBQS~1009-0252}
\def\btreize{B~1359+154}
\def\hevingt{HE~2149-2745}
\def\clover{H~1413+117}
\def\kmsmpc{km~s$^{-1}$~Mpc$^{-1}$}
\def\hyp{{\it Hyperz}}   
\def\ho{H$_0$ }
\def\sex{SExtractor}

  \title{A search for clusters and groups of galaxies on the line of sight towards 8 lensed quasars  \thanks{Based on observations obtained with VLT/ANTU at  ESO-Paranal Observatory (programs 67.A-0502 and 69.A-0247) and with the Hubble Space Telescope, operated by NASA.}}

  \subtitle{}
   
  \author{C. Faure\inst{1} \and D. Alloin\inst{2,4} \and J.P. Kneib \inst{3,5} \and F. Courbin\inst{6} }

\institute{Universidad Cat\'olica de Chile, Departamento de Astronomia y Astrofisica, Casilla 306, Santiago 22, Chile
          \and
          European Southern  Observatory, Alonso de Cordova 3107,
         Casilla   19001,    Santiago   19,   Chile    
         \and   
         Observatoire  Midi$-$Pyr\'en\'ees, UMR
         5572, 14  avenue Edouard  Belin, 31400 Toulouse,  France 
	 \and
	 URA2052CNRS/SAp/CEA, L'Orme des Merisiers, 91191 Gif sur Yvette, France
	 \and
         Caltech Astronomy  Department, Mail Code  105-24, Pasadena, CA
         91125, USA
	 \and
	 Institut
         d'Astrophysique et de  G\'eophysique, Facult\'e des Sciences,
         Universit\'e de Li\`ege, All\'ee du 6 ao\^ut, 17, B5C, Li\`ege 1,  
         Belgium 
}

\date{Accepted: 19/05/2004}

\authorrunning{C. Faure et al.}

\titlerunning{A search for clusters and groups of galaxies on the line of sight towards 8 lensed quasars}

\abstract{In this paper we present  new ESO/VLT FORS1 and ISAAC images of the fields around eight gravitationally lensed quasars:  CTQ~414, HE~0230-2130, LBQS~1009-0252, B~1030+074, HE~1104-1805,  B~1359+154, H~1413+117 and HE~2149-2745. When available and deep enough, HST/WFPC2 data were also  used to infer  the photometric redshifts of the galaxies around the quasars. The search of galaxy overdensities in space and redshift, as well as a weak-shear analysis and a mass reconstruction are presented in this paper. We find that there are most probably galaxy groups towards  CTQ~414, HE~0230-2130, B~1359+154,  H~1413+117 and HE~2149-2745, with a mass $\leq$4~10$^{14}$M$_\odot$h$^{-1}$. Considering its photometric redshift, the galaxy group discovered in the field around HE~1104-1805 is associated with the quasar rather than with the lensing potential.
}

 \maketitle
\section{Introduction}
Gravitational lensing of distant quasars is a powerful tool to address important cosmological and astrophysical questions. These include the distribution of dark matter in galaxies (Keeton, Kochanek \& Falco 1998), the determination of the Hubble parameter H$_0$ through the measurement of the time delay between the quasar image light-curves (Refsdal 1964), as well as ultimately the measurement of the value of the cosmological constant $\Lambda$ (Kochanek 1996). All the complexity in resolving these  questions is due to the difficulty  in modeling the lensing potential. Until now, most of the multiply imaged quasars were thought to be lensed by  one  main and single  lensing  galaxy, disregarding the local environment. Yet, while modeling these systems, it has been  realized that the addition of an "external shear" contribution is mandatory in many cases to reproduce the observations (Keeton, Kochanek \& Seljak 1997).  It is now known, that in at least one fourth of the  cases, the lensing potential is much more complex than a single lensing galaxy and that it is associated with a galaxy cluster or a galaxy group. Actually, the first X-ray detection of galaxy groups associated with the lensing potential in quadruple quasars has been  made recently towards the quasars RXJ~0911+0551 (Morgan et al. 2001), Q~0957+561 (Chartas et al. 2002), PG~1115+80 and B~1422+231 (Grant et al. 2003). Optically, many  galaxy clusters or groups have been discovered in the line of sight towards gravitationally lensed quasars, like PG~1115+80 (Kundic et al. 1997),  Q~0957+561 (Angonin-Willaine et al. 1994),  the {\it Cloverleaf} (Kneib et al. 1998b), RXJ~0911+0551 (Kneib et al. 2000) and  SBS~1520+530 (Faure et al. 2002). \\
In this paper we present the new VLT/FORS1 and ISAAC observations and wide-field analysis of eight  gravitationally lensed quasars in the southern  sky: \ctq, \hezero, \lbqs, \bdix, \heonze, \btreize, \clover\, and \hevingt. In previous studies of  these systems, the modeling of the lensing potential suffered from a lack of observational constraints and a poor knowledge of the quasar image surroundings. The modeling led indeed to the conclusion that the contribution of an external shear was  mandatory for most of these cases to reproduce the quasar image configuration using a realistic model for the main lensing galaxy. 
 In this paper, we  first provide  a brief summary of former  studies for each system (Section \ref{summary}). Then we describe the new ground-based VLT/UT1 observations, and the HST dataset  retrieved from the archive (Section \ref{data}).    In Section \ref{over}  we discuss  and give the results of our search for galaxy overdensities,  and in Section \ref{zphotsect} we provide the galaxy photometric redshifts. In Section \ref{weakmass}, we present the weak-shear analyses and the mass reconstructions  performed from the deep FORS1 images. And finally, in Section \ref{discussion}, we provide final remarks  and present our conclusions. 
We adopt $H_0$=65 km~s$^{-1}$~Mpc$^{-1}$, $\Omega$=0.3 and $\Lambda$=0.7
throughout the paper.


\section{Lens sample}\label{summary}

\begin{table*}
\renewcommand{\arraystretch}{1.0}
\centering
\begin{center}
\caption{ Properties of the lensed quasar sample. Column 1: quasar name. Columns 2, 3: right ascension and declination of quasar image A (J2000). Column 4: quasar redshift. Column 5: lensing galaxy redshift, where ``?'' means ``unknown'' and  ``(p)'' means ``photometric redshift''.  Column 6: size of the system corresponding to the maximal angular separation between the quasar images (in arcsecond). Column 7: system configuration. Column 8: predicted value of the external shear strength and position angle of the shear (1): Leh\'ar et al. 2002, (2): Saha \& Williams 2003, (3): Lopez et al. 1998, (4) Rusin et al. 2001. The symbol ``-'' means that the system has never been modeled.     }
\begin{tabular}{|c|l|l|l|l|l|l|l c|}
\hline
Quasar & RA & DEC & z$_{qso}$ & z$_{lens}$ & Size &  Configuration & Ext. shear and & PA(deg) \\
   
\hline
\ctq &01$^h$58$^{m}$41.43$^s$ & -43$^o$25\arcmin3.4\arcsec  &1.29 & ? &1.22\arcsec  &double &- & \\
\hline
\hezero&02$^h$32$^{m}$33.1$^s$&-21$^o$17\arcmin26\arcsec& 2.16 & $\leq$1.6 &2.10\arcsec   & quadruple&- &\\
\hline
\lbqs & 10$^h$12$^{m}$15.71$^s$&-03$^o$07\arcmin02\arcsec&2.74&0.88(p)&1.54\arcsec & double& 0.053$^1$& -36  \\
\hline
\bdix & 10$^h$33$^{m}$34.08$^s$ &+07$^o$11\arcmin25.5\arcsec&1.54 &0.599& 1.56\arcsec
 &double&0.09$^1$& -39   \\
\hline
\heonze& 11$^h$06$^{m}$33.45$^s$ &-18$^o$21\arcmin24.2\arcsec &2.32 & 0.73&3.19\arcsec& double& 0.05$^1$ & -38  \\
\hline
\btreize&14$^h$01$^{m}$35.55$^s$ & +15$^o$13\arcmin25.6\arcsec & 3.235 & ? &1.71\arcsec & sextuple& 0.12 to 0.26$^4$ & 71 to 97\\
\hline
\clover& 14$^h$15$^{m}$46.40$^s$ &+11$^o$29\arcmin41.4\arcsec &2.55& 0.9(p) &1.10\arcsec& quadruple & 0.08$^2$ & 70\\ 
\hline
\hevingt&21$^h$52$^{m}$07.44$^s$&  -27$^o$31\arcmin50.2\arcsec&2.03& 0.5(p)& 1.71\arcsec &double &  $>$0.21$^3$ & 58\\ 
\hline

\end{tabular}
\label{castles}
\end{center}
\end{table*}
The goal of this project is  to map the line of sight towards a large number of  lensed quasars in  the southern sky, up to the redshift of the lens.
Indeed, the systems studied in this paper have not yet been  modeled in a satisfactory manner due to the lack of constraints on the line of sight distribution towards the quasar on large scale.  In particular, little information is available in order to characterize the external shear due to possible intervening galaxy clusters or groups at high redshift along their line of sight. \\
In this section, we present a brief summary of the previous  studies performed on these systems. The quasar coordinates  and redshifts, as well as some characteristics of the lensing potential are given in Table \ref{castles} (from CfA-Arizona Space Telescope Lens  Survey \footnote{Gravitational Lens Database: http://cfa-www.harvard.edu/castles/},CASTLES; Mu\~noz et al. 1998). \\
As in most  papers about gravitationally lensed quasars, we will use the convention that the brightest quasar image (at the date and in the filter of the discovery)  is labeled A, the second bright one, is labeled B, etc.
\subsection{CTQ~414}
\ctq\, (also named Q~J0158-4325) has been first observed  in the Cal\'an-Tololo Survey (CTS; Maza et al. 1995), and was fully identified as a doubly imaged quasar at  redshift $z=$1.29  by Morgan et al. (1999). The quasar images are separated by 1.22\arcsec. A PSF subtraction of the quasar images reveals an extended object lying in between the quasar images: it is thought to be the lensing galaxy. It is located 0.72\arcsec\, West to image A. The presence of the lensing galaxy  has been confirmed by HST NICMOS observations (CASTLES). No mass model has been performed for this system until now.

\subsection{\hezero}
\hezero\,    has been discovered  in the course of the Hamburg/ESO survey (Wisotzki et al. 1996), and confirmed to be a lensed quasar by Wisotzki et al. (1999). It is a quadruply imaged quasar at $z=$2.162, with image angular separations from 0.7\arcsec\, to 2.1\arcsec. The lensing galaxy was also observed and its apparent magnitude suggests that it is  at redshift z$_l \leq$1.6. No mass model has been proposed so far.

\subsection{\lbqs}
\lbqs\,  was studied by Surdej et al. (1993) and  Hewett et al. (1994). It is a doubly imaged quasar at redshift $z=$2.74. The images are separated by 1.53\arcsec\, and show strong Mg II absorption lines at $z=$0.869. A second quasar, quasar C, at redshift $z=$1.62, lies by chance  at only 4.6\arcsec\, from quasar image A. The lensing galaxy was detected, close to quasar image  B, in the HST $I$- and $V$-band frames (Leh\`ar et al. 2000).\\
In the lens model of Leh\`ar et al. (2000), the  host galaxy of quasar C  is supposed to produce a shear of $\gamma\sim$0.053 (after correcting for redshift differences) and, according to these authors,  could  dominate the external shear when fitting the main lensing galaxy by a Singular Isothermal Ellipsoid (SIE).  A more recent model of the lens , requires an external shear: $\gamma$=0.017$\pm$0.009, with an orientation: PA$_{\gamma}$= 11$\pm$16~deg, if the lensing galaxy is modeled by a Singular Isothermal Sphere (SIS)(Claeskens et al. 2001).

\subsection{\bdix}
The bright radio source \bdix\,  is a two-component gravitationally lensed system  at $z=$1.535 (Xanthopoulos et al. 1998). It was discovered during the course of the Jodrell-Bank VLA Astrometric Survey (JVAS, Patnaik et al. 1992, Patnaik 1993, Browne et al. 1998, Wilkinson et al. 1998). The two quasar images are separated by 1.56\arcsec. The lensing galaxy was discovered by Fassnacht \& Cohen 1998, at a redshift z$_l$=0.599. It shows substructures that could be intrinsic  to the galaxy or that could trace in fact an  interacting galaxy system (Jackson et al. 2000).\\
A first model of the lensing potential was proposed by Xanthopoulos et al. (1998). They modeled the lensing galaxy by a SIE and a flux ratio between the quasar components of 20:1. They have estimated the mass of the lensing galaxy within the Einstein ring to be  1.3 $\times$ 10$^{11}$ M$\odot$. Then, Leh\`ar et al. (2000) proposed a two-component lens model,  the main lensing galaxy G being modeled by a SIE, and the second one, G\arcmin,  modeled by a SIS. They concluded that, without external shear, the orientation of the SIE is inconsistent with that observed for  the light of galaxy  G, whereas it provides a good fit of the lens constraints. This disagreement is still pending.

\subsection{\heonze}
\heonze\,  (Wisotzki  et al. 1993)  is a doubly imaged quasar at redshift $z=$2.319, with image  angular separation of 3.19\arcsec. In 1998, Courbin et al. discovered the lensing galaxy from ground-based near-IR ($J$- and $K$-band observations). Remy et al. (1998) have also detected the lensing galaxy in their WFPC2/$I$-band image, but not in their $V$-band image. Therefore,  they suggested that the lensing galaxy is  an early type galaxy. The redshift of the lensing galaxy  is z$_l$=0.729 (Lidman et al. 2000). 
The recent  time-delay measurement (Ofek \& Maoz 2003) gives a value of $\Delta t$=-161$\pm$7 days (68\% confidence level), in disagreement with former results by Gil-Merino et al. 2002 ($\Delta t$=-310$\pm$20 days) and  Pelt et al. 2002  ($\Delta t$=-255 days).
Moreover, this system is known to show conspicuous  microlensing effects (Wisotzki et al. 1993, Courbin et al. 2000, Gil-Merino et al. 2002, Schechter et al. 2003),  making it more difficult  to model. Several mass models have been performed (Wisotzki et al. 1998, Courbin et al. 2000, Schechter et al. 2003), all leading to the conclusion that the contribution of an external shear is mandatory to reproduce simultaneously the quasar image configuration, the time-delay and the flux ratios, using a realistic ellipticity and mass for the lensing galaxy.
  
\subsection{\btreize}
The complex lensed system \btreize\,  consists in a set of six images of a quasar at $z=$3.235 (Rusin et al. 2001). The main lensing potential is due to a group of three galaxies at z$\sim$1. Rusin et al. have assumed that the true  number of lensed quasar images is nine, and that only six of them are amplified and detected. They derived mass models, assuming four different galaxy potentials for the three main lensing galaxies: SIS, SIE, Pseudo-Jaffe spherical (PJS) and Pseudo-Jaffe ellipticals (PJE) (for the definitions, see Keeton \& Kochanek 1998). None of these models matches well  the flux ratios and the relative distances between the quasar images. The complexity of this system requires more observational constraints to be correctly modeled.   

\subsection{\clover}
The {\it Cloverleaf}  is a distant quasar at $z=$2.558 discovered by Hazard et al. (1984) and identified as a lensed quasar by Magain et al. (1988). \clover\, shows four images, with angular separations from 0.76\arcsec\, to 1.1\arcsec.  The lensing galaxy was  discovered by Kneib et al. (1998b) in the NICMOS-2/F160w observations. A galaxy overdensity was also discovered on the WFPC2/$I$- and $R$-band (filter  $I$: F814w, filter $R$: F702w) images, in the magnitude ranges 23$\leq$I$<$25 and 23$\leq$R$<$25 (Kneib et al. 1998a,b). According to Kneib et al. (1998a), the color of the galaxies building the overdensity suggest a redshift of $z=$0.9. Several models have been performed for the lensing object, using also constraints from the CO-map of the quasar images  (Kneib et al. 1998b), and the ones  which reproduce best the observations  are those taking into account the galaxy overdensity discovered by Kneib et al. (1998a,b).   
\subsection{\hevingt}
The doubly imaged quasar \hevingt\, at redshift z = 2.033, presents an angular separation of 1.71\arcsec\, (Wisotzki et al. 1998). The lensing galaxy, appearing after PSF subtraction of the quasar images, has an $R$-band magnitude  R=20.4.
The V-R color of the lensing galaxy, if elliptical, corresponds to a  maximum redshift of 0.5. The lensing galaxy lies at almost equal projected  distance from both quasar images, and these images fall on a direction coincident with the major axis of the lensing galaxy. Lopez et al. (1998) proposed a model for the lens. They used a singular isothermal elliptical mass distribution (SIEMD, Kassiola \& Kovner 1993). Due to the singular geometrical configuration of the quasar images, an external shear contribution is mandatory in the model. Because of the non-detection of any galaxy cluster in the field (magnitude limit: R$\sim$23), Lopez et al.  imposed the external shear $\gamma$ to be less than 0.21. Their best  models give a lens mass of : M=1.5(2.4)$\times$10$^{11}$M$\odot$ for a lensing galaxy at redshift z$_l$=0.3(0.5), corresponding to a mass to light ratio of $ \mathcal{M}/\mathcal{L}_{\rm R}$=5(2). A time delay between the quasar images light-curves has been measured: $\Delta t$=103$\pm$12 days (Burud et al. 2002).


\section{Observations - data reduction}\label{data}
\begin{table}
\renewcommand{\arraystretch}{1.0}
\centering
\begin{center}
\caption{FORS1 and ISAAC observations. Column 1: quasar name. Column 2: filter. Column 3: total exposure time in kilo-seconds. Column 4: field of view  used for the analysis.  }
\begin{tabular}{|c|l|l|l|l|}
\hline
Quasar & Band &  Seeing & Exp Time &Field \\
\hline
\ctq& R &0.50\arcsec &1.45 &6.5\arcmin$\times$6.6\arcmin \\
       & J      &0.51\arcsec &2.00 &2.3\arcmin$\times$2.4\arcmin  \\
       & Ks     &0.50\arcsec & 1.50&2.2\arcmin$\times$2.2\arcmin  \\
 \hline
\hezero & R& 0.74\arcsec & 3.40 & 6.4\arcmin$\times$6.6\arcmin  \\
            & J     & 0.58\arcsec & 2.60 & 2.5\arcmin$\times$2.4\arcmin  \\
           & Ks  & 0.50\arcsec & 2.70  & 2.1\arcmin$\times$2.4\arcmin \\
\hline
\lbqs & R &0.72\arcsec &3.90 & 6.5\arcmin$\times$6.6\arcmin \\
              &J      &0.44\arcsec &3.80& 2.4\arcmin$\times$2.5\arcmin \\
             &K       &0.53\arcsec & 4.50&2.2\arcmin$\times$2.2\arcmin\\
\hline
\bdix & R &0.48\arcsec &6.00&6.6\arcmin$\times$6.6\arcmin \\
      & J & 0.60\arcsec & 3.60 & 2.4\arcmin$\times$2.4\arcmin \\
      & Ks & 0.60\arcsec &4.80 & 2.3\arcmin$\times$2.3\arcmin\\
\hline
\heonze& R &  0.66\arcsec &0.80 &6.6\arcmin$\times$6.7\arcmin  \\
             & J     & 0.51\arcsec &3.50& 2.4\arcmin$\times$2.4\arcmin  \\
             & Ks     & 0.34\arcsec & 4.80 & 2.2\arcmin$\times$2.3\arcmin \\
\hline
\btreize &R  &0.62\arcsec&4.80&6.6\arcmin$\times$6.7\arcmin  \\
 &J &  0.64\arcsec & 3.60& 2.3\arcmin$\times$2.3\arcmin\\
 &Ks&  0.47\arcsec & 4.50& 2.4\arcmin$\times$2.4\arcmin\\
\hline
\clover&R&0.58\arcsec & 1.70&6.2\arcmin$\times$6.2\arcmin\\
      &J  &0.59\arcsec& 2.60& 2.6\arcmin$\times$2.6\arcmin  \\
     &Ks &0.49\arcsec&  3.60&  2.6\arcmin$\times$2.6\arcmin  \\
\hline
\hevingt&R&0.57\arcsec &1.70 &6.5\arcmin$\times$6.7\arcmin \\
            &J    &0.48\arcsec& 2.60& 2.3\arcmin$\times$2.5\arcmin \\
            &K    & 0.47\arcsec & 3.60&2.3\arcmin$\times$2.5\arcmin \\
\hline

\end{tabular}
\label{obs}
\end{center}
\end{table}
\subsection{Ground-based dataset}
\begin{figure*}[!p]
\begin{center}
\includegraphics[width=15cm]{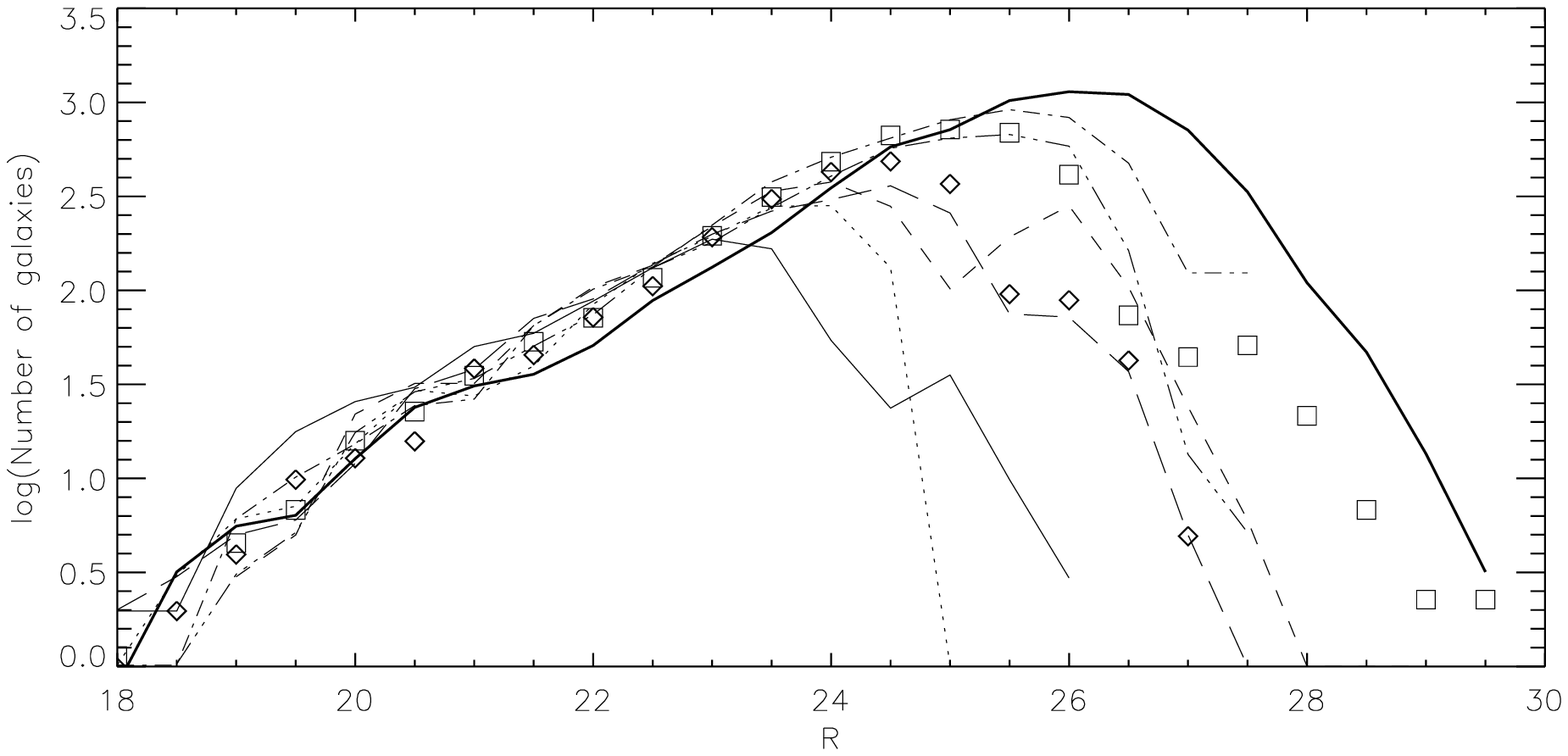}
\includegraphics[width=15cm]{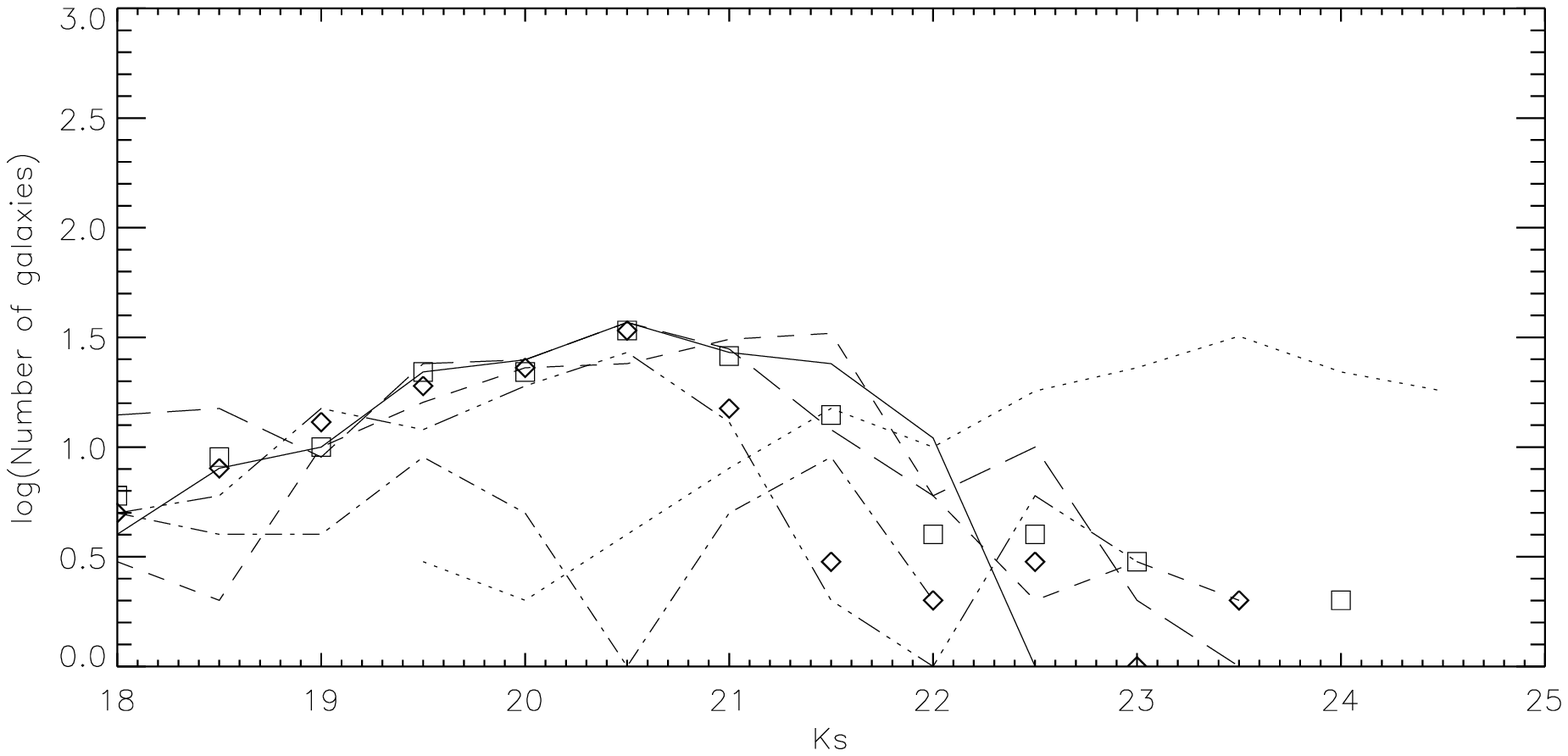}
\includegraphics[width=15cm]{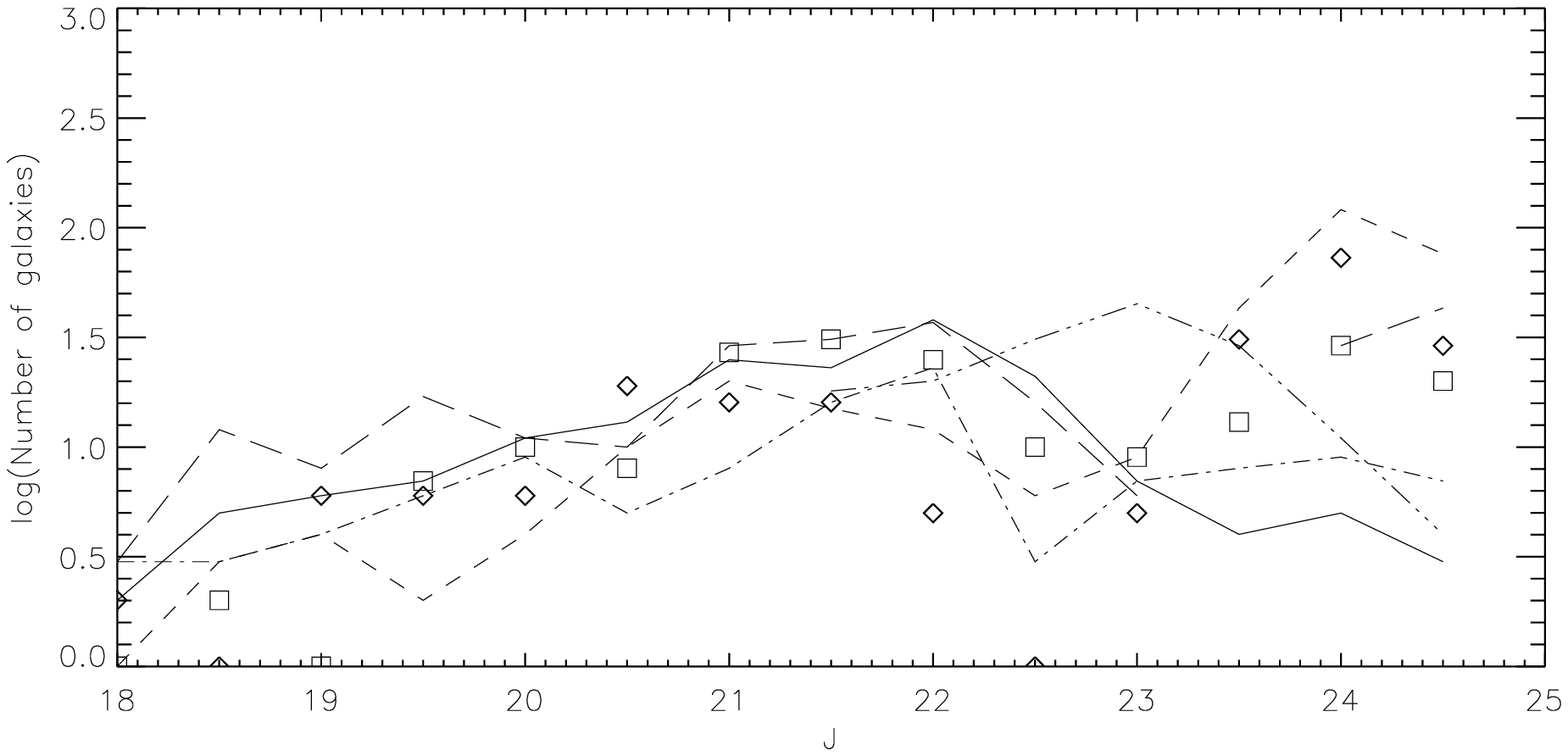}
\caption{\label{num} Logarithm of the number of galaxies as a function of magnitude. Top panel: in the FORS1 ($R$-band) fields around the lensed quasars and in the FORS Deep Field (FDF) Survey (thick solid line; Heidt et al. 2003). All the plots are rescaled to a 6.6\arcmin$\times$6.6\arcmin\, field of view. Middle panel: in the ISAAC ($Ks$-band) fields around the lensed quasars. Bottom panel: in the ISAAC ($J$-band) fields around the lensed quasars. Diamonds: \btreize; squares:  \clover;  long-dash line: \hevingt; solid line: \heonze; dotted line:  \lbqs; dash line: \bdix; dash-dot line: \ctq; dash-dot-dot-dot line: \hezero.}
\end{center}
\end{figure*}

 We took deep and high resolution FORS1 $R$-band images in order to exploit wide field images in the  analysis of   the lensing potential at large scale, performing a galaxy overdensity search and a weak-shear analysis. The near-IR images taken with the ISAAC instrument where mandatory to trace the stellar mass of galaxies at high redshift (z$\sim$3-4), and to measure galaxy  photometric redshifts from  multi-color analysis. \\

The ground-based observations consist in two datasets  taken with ISAAC (Cuby et al. 2002) and with FORS1 (Szeifert 2002) attached to the 8.2~m telescope VLT/ANTU, at ESO/Paranal. The imaging datasets were obtained during  ESO periods 67 and 69 (programs 67.A-0502(A,B,C) and 69.A-0247(A,B,C)).
 The ISAAC images  (1024 $\times$ 1024~pixels, 0.1484\arcsec\, per pixel) were taken in two near-IR filters: the $J$- and $Ks$-filters. We reduced the data following  standard procedures (ISAAC Data Reduction Guide, Amico et al. 2002). All the images ($J$- and $Ks$-bands) taken in P67 were affected by the ``odd-even column effect'', arising from an ISAAC mis-functioning during this period (this concerns objects \ctq, \hezero, \clover\, and \hevingt). This artifact was successfully removed through a Fourier analysis procedure (see ISAAC Handbook, Cuby et al. 2002).   When images of a same object were taken during different nights we have reduced all images to the same "seeing", through convolution by a Gaussian, before combining them. A summary of the observations is given in Table \ref{obs}.\\
 The FORS1  $R$-band images (2040 $\times$ 2040~pixels, 0.2\arcsec\, per pixel) were reduced using sky flat-fields and standard IRAF procedures for optical image reduction.   Saturated stars  were masked in order to perform the sky subtraction and the  galaxy extraction (done both  using SExtractor; Bertin \& Arnouts 1996)  . Quantitative information on the  quality of the images is provided in Table \ref{obs}.\\ 
The sizes and positions of the ISAAC fields relative to the  FORS1 fields are shown on Fig.\ref{iso1} and Fig.\ref{iso2}. \\
As shown in Table \ref{obs}, the total  exposure time varied from one system to the other, a consequence of the observations being conducted in service mode. In particular, some systems are under-exposed like \heonze\, in the $R$-band and  \ctq\, in the $Ks$-band.   
\subsection{HST imaging}
\begin{table}
\renewcommand{\arraystretch}{1.0}
\centering
\begin{center}
\caption{WFPC2 observations. Column 1: quasar name. Column 2: filter. Column 3: proposal P.I. and proposal ID. Column 4: total exposure time  (in ks).  }
\begin{tabular}{|c|l|l|l|}
\hline
Quasar & Band &  P.I and I.D.& Exp Time \\
      
\hline
\ctq&F555w  & Falco (9133)   &1 .60 \\
  & F814w& Impey (8268) & 2.00\\
 \hline
\heonze& F814w &Surdej (5958) & 1.00\\
\hline
\btreize&F555w  & Impey (8268) &5.20 \\ 
&F814w & Impey (8268) & 5.00\\
\hline
\clover& F814w &Turnshek (5621)& 5.20\\ 
\hline
\hevingt&F555w & Falco (8175)  & 1.60\\
 & F814w & Falco (7495) &1.60\\ 
\hline

\end{tabular}
\label{obsHST}
\end{center}
\end{table}

We have searched the HST public  archive for extra images of the quasar fields around: \ctq, \hezero, \heonze, \btreize, \clover\, and \hevingt, in the $I$- and $V$-bands (F814w and F555w). Details of the observations are given in Table \ref{obsHST}. 
In most cases we are able to use directly the WFPC2 association frames\footnote{ http://archive.eso.org/wdb/wdb/hst/asn\_wfpc2/form}, otherwise we used standard reduction procedures from the IRAF/STSDAS package. To project the HST images onto the sky we used the SWarp package from Bertin\footnote{http://terapix.iap.fr/complement/oldSite/soft/swarp/}. 


\section{Looking for galaxy overdensities}\label{over}
\begin{figure*}[!tp]
\begin{center}
\caption{\label{isoK} ISAAC $Ks$-band images of \hezero\, (left) and \clover\,(right). The contours (from 2$\sigma$, increasing by 1$\sigma$ steps) outline regions of same galaxy number density (label-value attached to each contour in  galaxies per arcmin$^2$). North is to the top, East  to the left. For {\bf \hezero}, the galaxy overdensity is located $\Delta\alpha$=40\arcsec,$\Delta\delta$=-3\arcsec\, relative  to quasar image A (coordinates (0,0)) and occurs in the magnitude range 19$\leq$Ks$<$22. For {\bf \clover},the galaxy overdensity is located $\Delta\alpha$=-3\arcsec\,/$\Delta\delta$=-6\arcsec\, relative  to  quasar image A (coordinates (0,0))  in the same magnitude range as for \hezero.   }
\end{center}
\end{figure*}
\subsection{Photometry and galaxy-star separation}
We have searched for galaxy number overdensities in the field around the quasar images. Indeed, if these overdensities are tracing galaxy clusters or galaxy groups, they must take part in the total lensing potential. 

\begin{figure*}[!p]
\begin{center}
\caption{\label{iso1}  FORS1 $R$-band images of \clover, \btreize, \bdix\, and \ctq. North is to the top, East to the left. The contours (from 2$\sigma$, increasing by 1$\sigma$) outline regions of same galaxy number density (label-value attached to each contour in galaxies per arcmin$^2$). The dash-line squares represent the position and relative size of the ISAAC field, while the solid-line squares represent the position, size and orientation of the WFPC2 field (in the case of \ctq\,, the  dash-point squares represent  the position, size and orientation of the WFPC2 $V$-band field, different from the one of the WFPC2 $I$-band ones). For each system,  quasar image A is located at  coordinates (0,0). For {\bf \clover}, 2 galaxies overdensities are discovered: at $\Delta\alpha$=-10\arcsec,$\Delta\delta=$-10\arcsec\, in the magnitude range 23$\leq$R$<$25, solid-line contours and  at $\Delta\alpha$=35\arcsec,$\Delta\delta=$-5\arcsec\, in the magnitude range 25$\leq$R$<$28, dash-line contours. For {\bf \btreize}, a  galaxy overdensity is found at $\Delta\alpha$=-8\arcsec,$\Delta\delta=$3\arcsec\, (22$\leq$R$<$23). For {\bf \bdix}, also one galaxy overdensity is detected ($\Delta\alpha$=-90\arcsec,$\Delta\delta=$-70\arcsec, 21$\leq$R$<$23). Finally, for {\bf \ctq}, a galaxy overdensity is discovered in the magnitude 24$\leq$R$<$25 at $\Delta\alpha$=-10\arcsec,$\Delta\delta=$-5\arcsec.}
\end{center}
\end{figure*}

\begin{table*}
\renewcommand{\arraystretch}{1.0}
\centering
\begin{center}
\caption{Characteristics of the galaxy number overdensities. Column 1: quasar name. Column 2: limit for the galaxy catalog  (said to be  as complete as the  FDF galaxy catalog   where the curves in Fig.\ref{num}-top  do join). Column 3: magnitude range over which  we can study the significance of the galaxy overdensity. Column 4: position of the center of the galaxy number overdensity relative to quasar image A (see Figs \ref{iso1} and \ref{iso2}). The uncertainty on the exact position of the galaxy overdensity is about 5\arcsec. Column 5: galaxy density at the location of the  galaxy overdensity center. Column 6: FDF mean galaxy density in the magnitude range given in column 3. Column 7: FDF maximal galaxy density in the same magnitude range. Column 8:  detection level (in $\sigma$) of the cluster compared to the FDF mean galaxy density level. Column 9: detection level of the cluster (in the magnitude range given in parentheses) above  the galaxy background level measured in the same frame.     }
\begin{tabular}{|c|c|c|c|r|r|r|r|r|r|}
\hline
 & Compl.       &Mag        &Galaxy overdensity       & Density          & FDF       & FDF        &  Above  & Above\\
 & limit       &range       &   location    &             & mean      & max        & FDF  &    background \\
&             &            &   $\Delta\alpha$,$\Delta\delta$   &  gal/(\arcmin)$^2$ &   density  & density           & density        &   mean                \\
\hline
\ctq & 25.2& 24$\leq$R$<$25&     $-$10\arcsec,$-$05\arcsec     & 29.0 & 14.4 & 25.0 &2.6$\sigma$& 4$\sigma$ (24$\leq$R$<$25)\\
\hline
\hezero &  24.5& 24$\leq$R$<$24.5 & $+$40\arcsec,$-$05\arcsec& 7.7  & 5.4& 14.8  & 1.1$\sigma$& 2.5$\sigma$ (24$\leq$R$<$25)\\
 \hline
\lbqs & 24.0 &23$\leq$R$<$24 & $+$60\arcsec,$+$40\arcsec&10.8  & 5.2& 10.0 & 2.8$\sigma$& 3.5$\sigma$ (23$\leq$R$<$25) \\
\hline
\bdix & 24.0 & 21$\leq$R$<$23 &$-$90\arcsec,$-$70\arcsec& 11.0& 3.2 & 7.0 & 5.6$\sigma$&4.5$\sigma$ (21$\leq$R$<$23) \\
\hline
\heonze& 23.3 & 23$\leq$R$<$23.3&$~~$00\arcsec,$+$50\arcsec& 9.0 &1.2 &2.0  & 26$\sigma$ & 4$\sigma$ (23$\leq$R$<$24)  \\
\hline
\btreize & 24.3& 22$\leq$R$<$23 &$-$08\arcsec,$+$03\arcsec&4.6 & 2.2 & 6.0  & 3.0$\sigma$ & 5$\sigma$ (22$\leq$R$<$23) \\
\hline
\clover& 25.0 &23$\leq$R$<$25&$-$10\arcsec,$-$10\arcsec& 38.3 & 19.7 & 32.5 &  2.3$\sigma$ &5$\sigma$ (23$\leq$R$<$25)\\
\hline
\hevingt & 24.0 & 23$\leq$R$<$24&$+$08\arcsec,$+$07\arcsec& 10.6& 5.2 &10.0&  2.6$\sigma$ & 6$\sigma$ (23$\leq$R$<$25) \\
\hline
\end{tabular}
\label{density}
\end{center}
\end{table*}

The extraction and photometry of the objects from  all frames were performed using the {\it SExtractor} software (Bertin \& Arnouts 1996).
Galaxies are separated from  stars using a combination of FWHM (Full Width-Half Maximum)  versus magnitude (MAG\_BEST in SExtractor) analysis and $\mu_{max}$ (peak surface brightness in mag~arcsec$^{-2}$) versus magnitude analysis (see Bardeau et al. 2004). The stars are the brightest objects of the catalogs, having the highest peak surface brightness and the smallest FWHM.

\begin{figure*}[!p]
\begin{center}
\caption{\label{iso2}FORS1 R-band images of \hezero, \heonze, \lbqs\, and \hevingt. The contours (from 2$\sigma$, increasing by 1$\sigma$) outline regions of same galaxy number density (label-value attached to each contour in  galaxies per arcmin$^2$). The dash-line squares represent the position and relative size of the ISAAC field, while the solid-line squares represent the position, size and orientation of the WFPC2 field. North is to the top, East  to the left. For {\bf \hezero}, one galaxy overdensity is  discovered at $\Delta\alpha$=40\arcsec,$\Delta\delta=$-5\arcsec\, (24$\leq$R$<$25). For {\bf \heonze}, two galaxy overdensities are discovered: at  $\Delta\alpha$=0\arcsec,$\Delta\delta=$50\arcsec, 23$\leq$R$<$24, solid-line contours   and at  $\Delta\alpha$=60\arcsec,$\Delta\delta=$-50\arcsec, 24$\leq$R$<$25, dash-line contours. For {\bf \lbqs}, a  galaxy overdensity is found at  $\Delta\alpha$=60\arcsec,$\Delta\delta=$40\arcsec\, to quasar image A (23$\leq$R$<$25). For {\bf \hevingt}, the closest  galaxy overdensity to the quasar images is centered   at  $\Delta\alpha$=8\arcsec,$\Delta\delta=$-7\arcsec (23$\leq$R$ <$25). }
\end{center}
\end{figure*}

\subsection{Overdensities above the galaxy background level}\label{42}
To search for galaxy number overdensities, we have proceeded as follow: 
for each field, the  galaxy magnitude  catalogs were split in  magnitude slices of one magnitude  width, allowing  a first-order  study of different redshift slices. For each magnitude slice, we have cut the image in a grid of 10$\times$10 cells \footnote{$\sim$ 40\arcsec\, size  cell in the $R$-band image, 15\arcsec\, in the $J$- and $Ks$-images}. For each galaxy $G_1$ in a cell we have measured  the distance between $G_1$ and its ninth closest galaxy $G_{10}$.  Then, we have  calculated the area, $\mathcal{A}$, covered by a circle of radius $\mathcal{R}$=$ | G_1-G_{10} | $. By definition, the number of galaxies included in the area  $\mathcal{A}$, with radius $\mathcal{R}$=$ | G_1-G_{10} | $, is 10. Therefore, the galaxy number  density associated to the galaxy  $G_1$ is the ratio 10/$\mathcal{A}$ . A mean galaxy density for each  cell is then derived,  averaging the densities associated to all galaxies in  the cell.  The grid is chosen considering that the central part of a galaxy cluster  would  cover between 1 and 4 cells, depending on its redshift and on  its central density. Then, the isodensity contours of the number of galaxies per arcmin$^2$ across the field are drawn: when an overdensity is observed in two consecutive magnitude slices, the two sub-catalogs are added in order to increase the signal to noise of the overdensity detection. The results are displayed in Fig. \ref{iso1} and Fig. \ref{iso2}.\\

To evaluate the significance of these detections we have performed the following tests.
As  mentioned above, a galaxy cluster should appear in  one to four contiguous over-densed cells in comparison to the rest of the frame. We have plotted the number of cells versus the galaxy density in the cells. This plot can be fitted by a Gaussian (see Faure et al. 2003 for a similar analysis), providing the mean galaxy  density in a field for a given magnitude range, as well as a standard deviation $\sigma$ (Fig. \ref{iso1} and Fig. \ref{iso2}). 
This estimator of detection is given in column 9 in Table \ref{density} and results are discussed in Section \ref{resiso}.\\

\subsection{Comparison with the FORS Deep Field}

For the $R$-band images we   have also  used the FORS Deep Field\footnote{The FDF photometric catalog is available from:  http://cdsweb.u-strasbg.fr/cgi-bin/qcat?J/A+A/398/49.} (FDF; Heidt et al. 2003) $R$-band galaxy catalog as a reference field. The $R$-band data set of the  FDF covers an area of $\sim$7.4\arcmin$\times$7.4\arcmin (26.4 kilo-second exposure time), up to  R=26.7 (50\% completeness). The FDF  field is  centered close to the South Galactic Pole (1$^h$6$^m$3.6$^s$, -25$^{\rm o}$45\arcmin46\arcsec, J2000). It has been chosen because there was {\it a priori} no previously known galaxy cluster in this field. \\

 The FDF dataset is much deeper than our dataset (see Fig. \ref{num}). So we are allowed to compare them up to the magnitude corresponding to the 100\% completeness limit of our dataset. These magnitudes are displayed in Table \ref{density}.\\

We have proceeded as follows:
 first, we have split the FDF catalog in magnitude ranges identical  to those where galaxy overdensities were discovered in the fields around the quasars. When an overdensity was detected in a magnitude range such that the upper magnitude  was higher than the completeness magnitude, we have compared only the part of the catalog for which  the upper magnitude was equal to the completeness magnitude (see Table \ref{density}, columns 2 and 3). 
We have partitioned the FDF in a grid of 10$\times$ 10 cells. For each cell, a galaxy density in a given magnitude range is measured and as detailed in Section \ref{42}, we have  fitted a Gaussian to the histogram describing the  ``number of cells'' versus  the ``galaxy density''. This  provides  the mean galaxy density, the standard deviation and the maximal density expected in a cluster-free region, in each magnitude range. The results of the comparison between the FDF and our dataset are displayed in Table \ref{density}.\\

\subsection{Results of the search for  galaxy overdensities}\label{resiso}

The galaxy  overdensities detected towards \ctq,  \bdix,  \heonze\, and \clover\, are well above the FDF mean level, revealing a 2 dimensional structure in the line of sight towards these quasars. On the contrary, the galaxy overdensities discovered towards \hezero,  \lbqs, \btreize\, and   \hevingt\, show values lower  than (or very close to) the maximal density measured in the FDF field. This could mean that these densities are compatible with the variance of the counts within the field of view. Hence, they  may correspond to areas where the count fluctuations reach a slightly higher level, still compatible with random deviation from the mean density. Another possibility is that there are undetected galaxy groups in the FDF, leading to values of the  galaxy densities  as strong as in the lens fields. \\
Comparing the detected overdensity to the mean galaxy density measured over the rest of the frame, we see that, apart from  the galaxy overdensities towards \hezero\, and  \lbqs\, (respectively detected at 2.5$\sigma$ and 3.5$\sigma$ above the background level), the overdensities measured towards the other six lensed quasars are all  significant (4$\sigma$ to 6$\sigma$). Notice that the high detection level (26$\sigma$) of the cluster towards \heonze\, compared to the FDF galaxy number density  is most  probably overestimated. This results from the fact that the galaxy densities are analyzed over a very narrow magnitude range  ($\Delta$m=0.3), and hence, a small galaxy sample ($\sim$100 galaxies), leading consequently to possibly large statistical effects.\\
Furthermore, there are two cases for which we have detected  more than one cluster in the line of sight towards the quasar images. Towards 
  \clover\, a second galaxy overdensity has been detected  in the magnitude range  25$\leq$R$<$28,  at  3$\sigma$ above the background level (centered 36\arcsec\, South-West to the quasar images). In a similar way, we find a second overdensity towards \heonze\, (78\arcsec South-West) in the magnitude range  24$\leq$R$<$25, at 5$\sigma$ above the background level. But in these two cases, the data used for the estimation extend below the  catalog completeness limit, and therefore we cannot assure whether these  are real detections or whether they correspond to  standard deviations at that level.\\

In most  cases, the number of galaxies detected in the near-IR frames is too low to perform a secure iso-density analysis. Nevertheless, in the $Ks$-band images towards \hezero\, and \clover\, we find again the galaxy overdensities seen in the $R$-band images (see Fig. \ref{isoK}).  The shifts between the positions of the overdensities in the $R$- and $Ks$- band images come mainly from the use of  different cell sizes and locations. \\

According to these results we conclude that, apart from the marginal galaxy overdensity towards \lbqs, all others are possible 3D-structures. That is what we are going to check in the following sections.

\section{Photometric redshifts and color-magnitude analysis}\label{zphotsect}


We can test whether the galaxy overdensities discovered  in Section \ref{over} are associated to galaxy clusters or galaxy groups by measuring the galaxy photometric redshifts. The redshift will also tell us whether these structures lie in  the line of sight  towards the quasar or whether they are at the quasar redshift.  
\subsection{Photometric redshifts }

Whenever a galaxy overdensity was  discovered on the line of sight towards the quasar images, we performed the photometric redshift analysis using  the {\it Hyperz} software (Bolzonella et al. 2000) and the full dataset available (described in Section \ref{data}). The input to {\it  Hyperz }  are the galaxy magnitudes  in {\bf at least} 4 filters \footnote{ A study of the accuracy of the code, according to the number of filters used to infer the redshifts, is detailed in Bolzonella et al. (2000)}. \hyp\, performs  a comparison between the photometric spectral energy distribution (SED) of the observed galaxy and those derived from a set of reference template spectra, using the same photometric system.
Reddening is  taken into account using  the Calzetti law (Calzetti et al. 2000). We estimate that the current dataset is deep enough to reach giant elliptical galaxies at  redshift up  to $z=$3.5. Two parameters are important to test the significance of the photometric redshifts found: the $\chi^2$ value of  the SED fit, and the probability of obtaining this $\chi^2$: P($\chi^2$). In  the following analysis, we only take into account the redshifts  obtained with $\chi^2\sim$1 and  P($\chi^2$)$>$70\%, considering that this is the limit of confidence for the redshift determination. \\


\begin{figure*}
\begin{center}
\includegraphics[width=6cm]{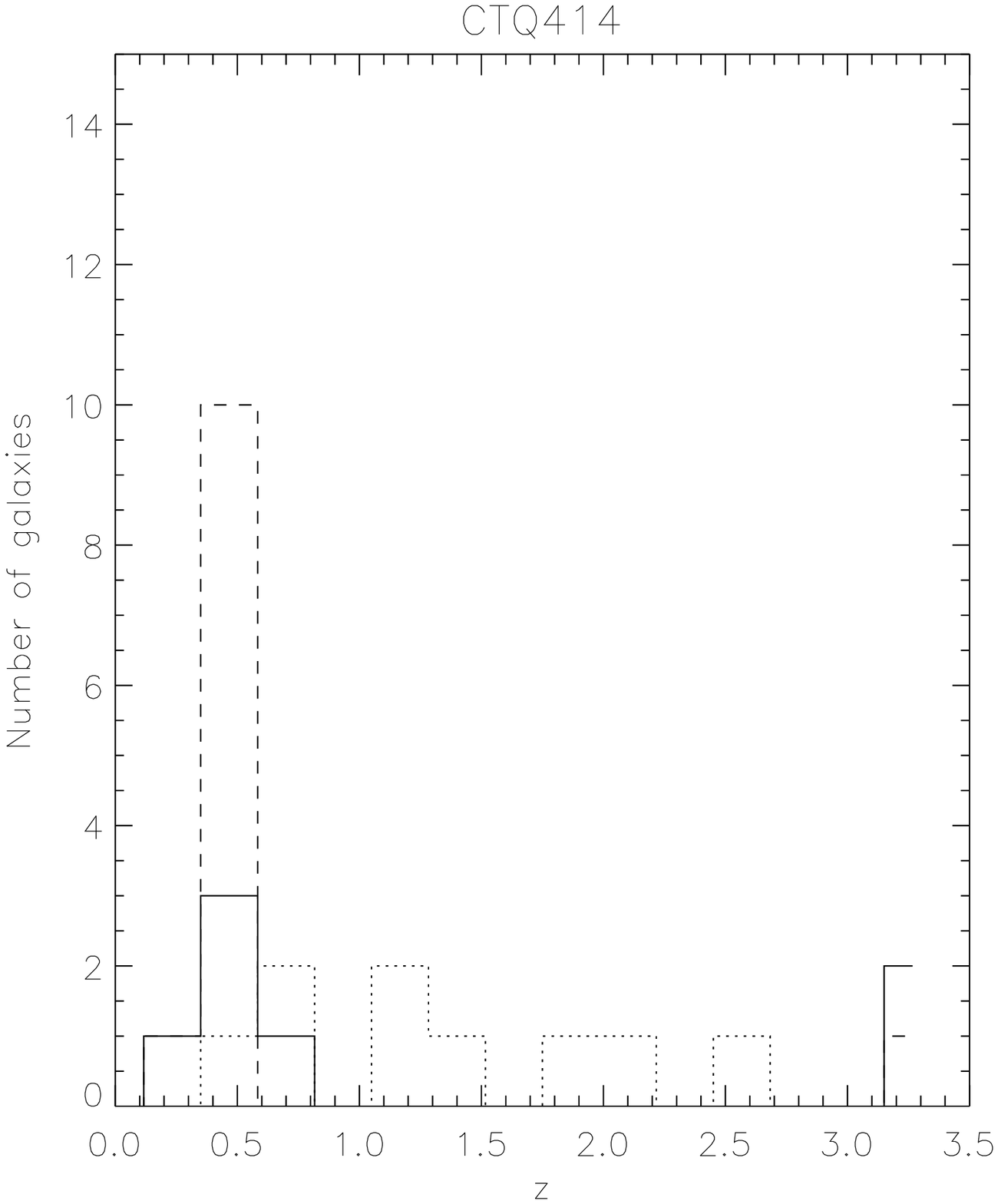}\includegraphics[width=6cm]{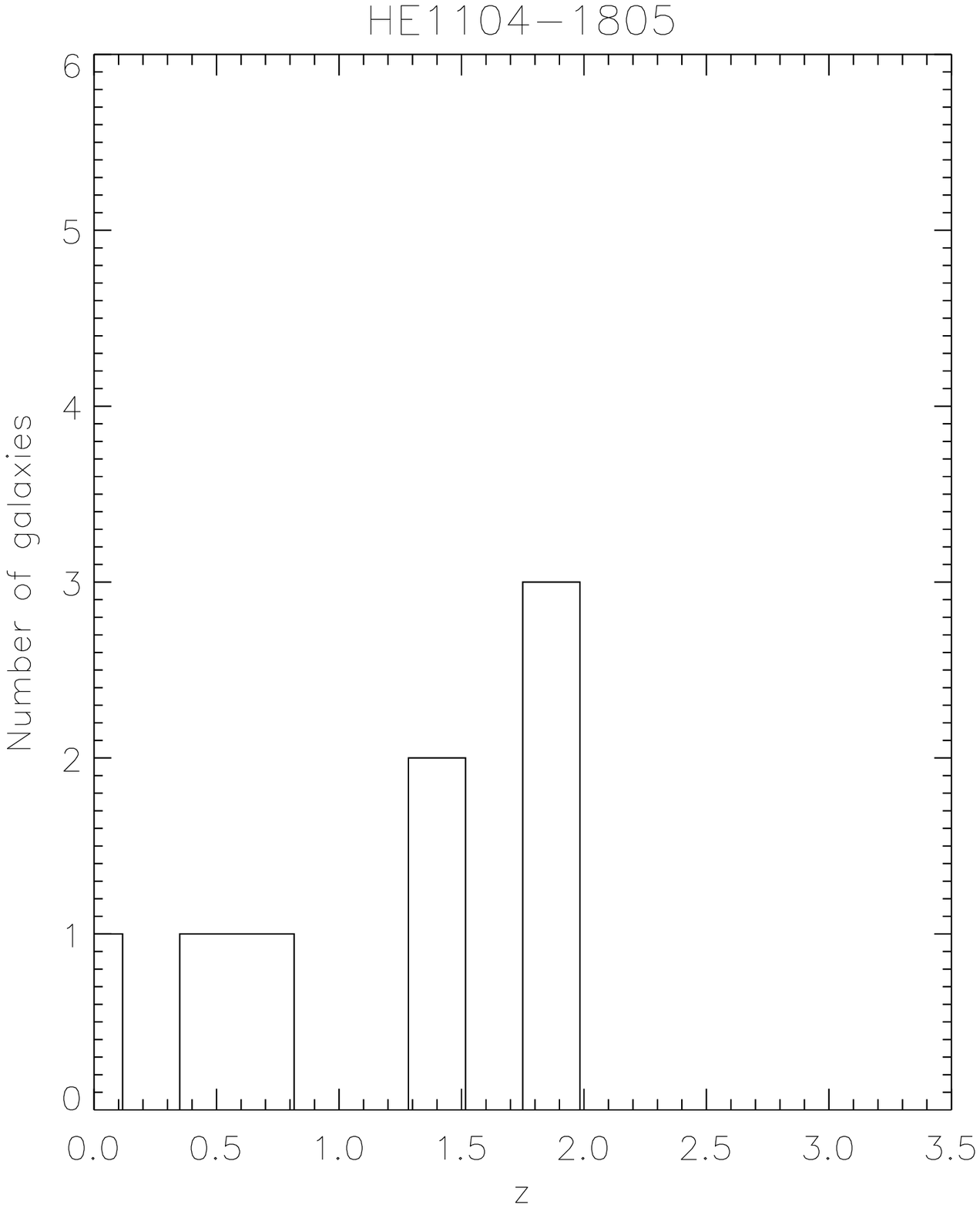}\includegraphics[width=6cm]{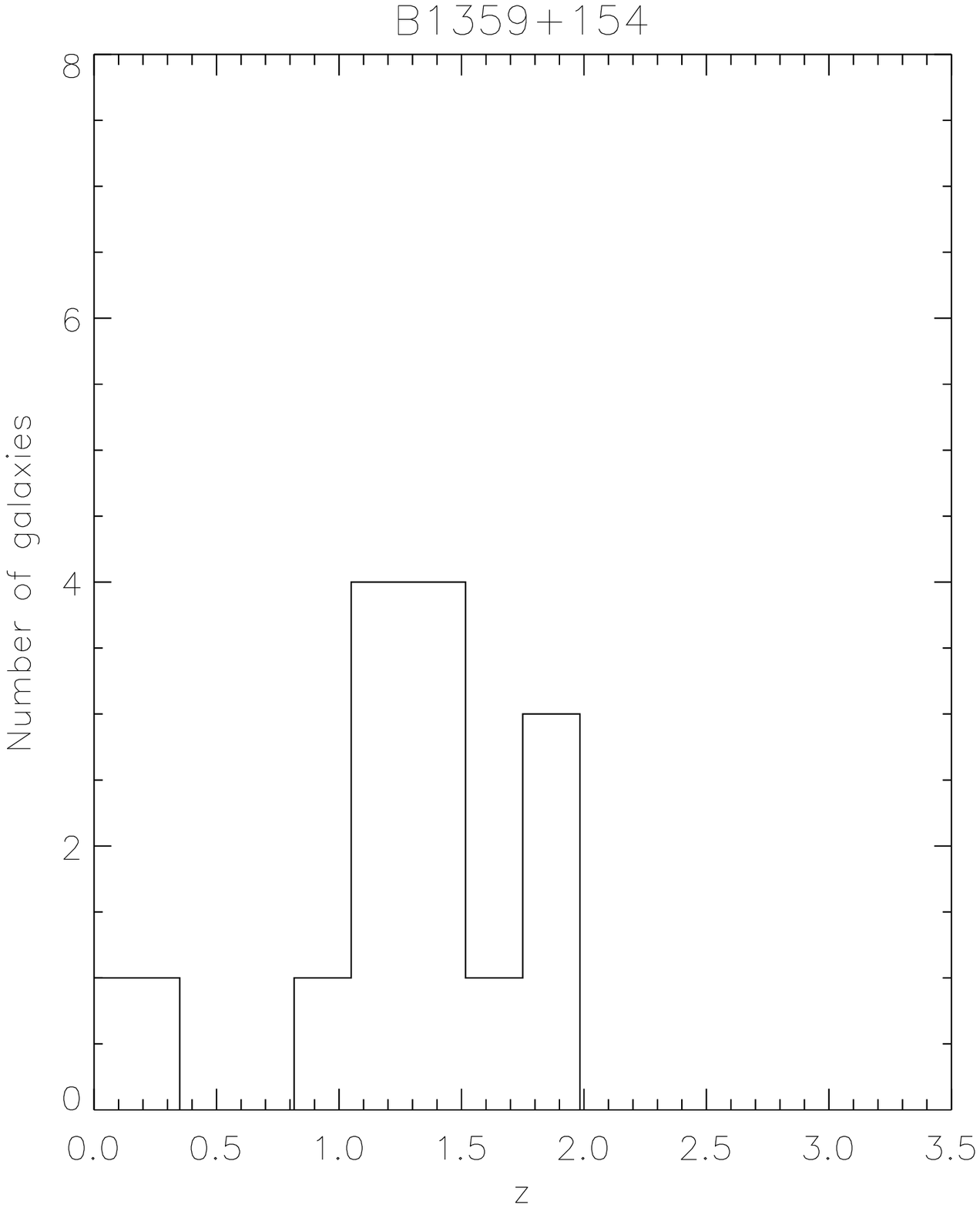}
\includegraphics[width=6cm]{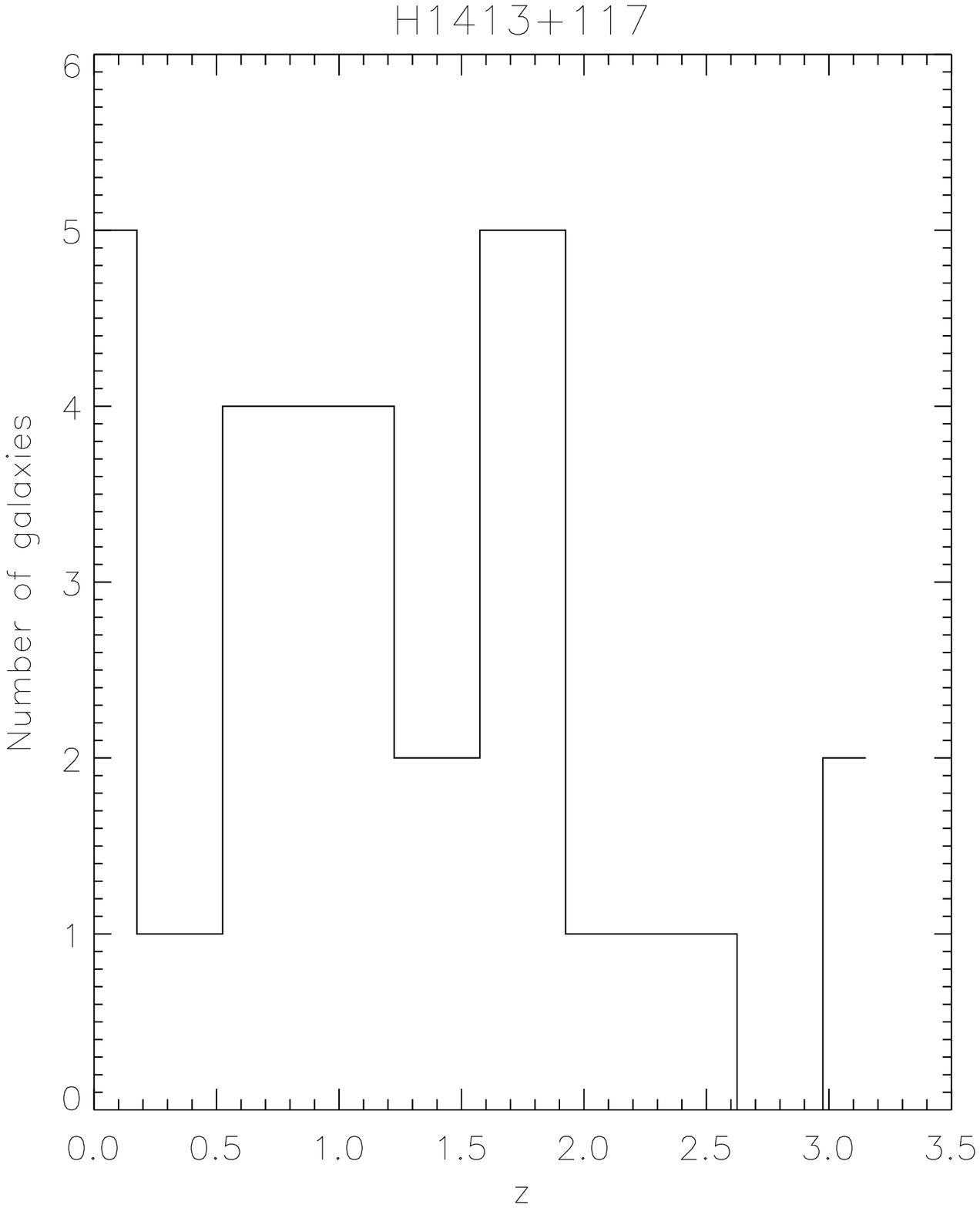}\includegraphics[width=6cm]{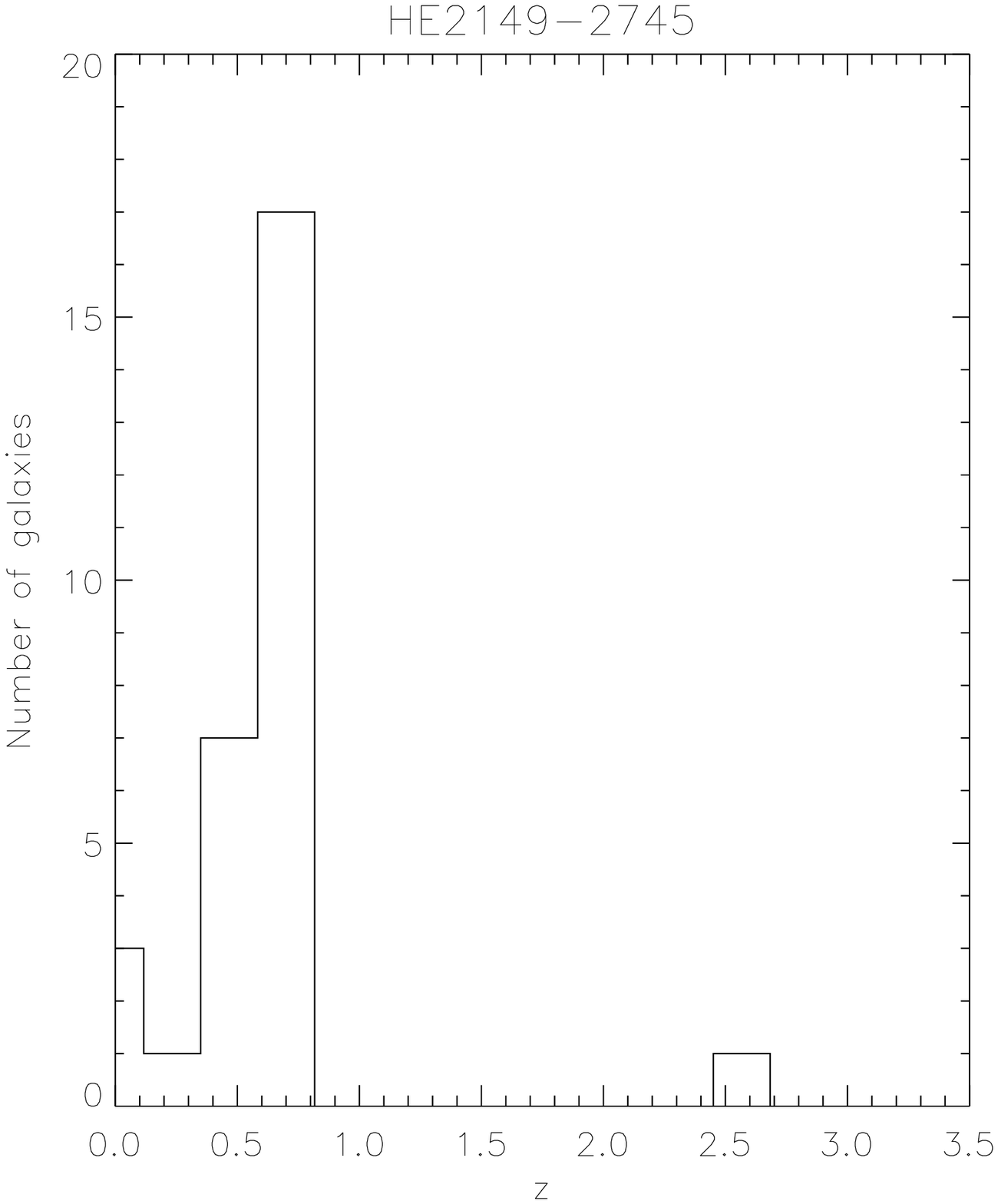}
\caption{\label{histo}  Number of galaxies versus photometric redshifts for {\bf \ctq\,} (top left hand panel), for {\bf \heonze\,} (top central panel), for {\bf \btreize\,} (top right hand panel), for {\bf \clover\,} (bottom left hand panel) and for {\bf \hevingt\,} (bottom right hand panel). For \ctq, the solid-line corresponds to  the photometric redshift distribution determined from the: V, R, I, J and Ks galaxy magnitudes; the dotted-line corresponds to the redshift determination using the: R, I, J and Ks galaxy magnitudes; the dash-line corresponds to the redshift determination using the:  V, R, J and Ks galaxy magnitudes. }
\end{center}
\end{figure*}

In the FORS1 field around  {\bf \ctq\,}
we have computed the galaxy photometric redshifts using the VLT and  HST data. The overlapping field  covers only  a very small area (see Fig. \ref{iso1}), hence only 7 galaxies have their redshift determined using 5 magnitudes (V, R, I, J and Ks). In addition, we tentatively  determined the redshifts of 12 galaxies from their  $V$-, $R$-, $J$- and $Ks$-band magnitudes, and of 9 galaxies from their $R$-, $I$-, $J$- and  $Ks$-band magnitudes. The results are plotted in Fig. \ref{histo}, showing a significant peak in redshift at $z=$0.5$\pm$0.1. In the following, this is assumed to be the redshift of the galaxy overdensity discovered towards the quasar.\\
 
For {\bf \hezero}, we have not been able to determine the  photometric redshifts of the galaxies building the observed marginal overdensity, because of the low depth of the WFPC2 images available for  this field.  However, according to   Wisotzki et al. (1999) the lensing galaxy redshift    is z$_l<$1.6. Assuming that the  lensing galaxy is  part of the galaxy cluster tentatively discovered,  we will use arbitrarily $z=$1.0 for the lensing potential redshift in the mass reconstruction.  \\

For {\bf \lbqs\,}  there is no significant galaxy overdensity  in the field (Section \ref{over}) and with the galaxy photometry available in only 3 bands ($R$, $J$ and $Ks$) we cannot  perform a photometric redshift analysis. Therefore, a weak-lensing analysis is necessary to define whether or not there is a dark source of shear generating the external-shear value predicted by Lehar et al. (2000, see Table \ref{castles}).     \\
 
For {\bf  \bdix}, the lack of multi-wavelength data makes impossible  photometric redshift measurements.   Moreover, the  galaxy overdensity  detected in the FORS1 field around \bdix\, (see Section \ref{over}) is centered outside the ISAAC field. Would this galaxy overdensity be a cluster or a group, it would be  too far away from the quasar images ($\sim$2\arcmin) to play any significant role in the lensing potential (shear strength negligible). Hence, as  for \lbqs, we expect the weak-shear analysis to conclude whether or not there is a galaxy group or cluster in the line of sight towards the quasar.    \\

Around  {\bf \heonze\,} we have derived the photometric redshifts of the galaxies present in the overlapping  HST and VLT fields (see Fig. \ref{iso2}). One WFPC2 chip covers mainly the galaxy overdensity located 50\arcsec\, North to the quasar images. The four galaxies detected both on this chip and on  the three VLT images have a redshift  in the range: 1.95$\leq$z$\leq$2.10. Therefore, this cluster is most probably associated with the quasar (z=2.33) rather than with the lens. The photometric redshift of the second galaxy overdensity has not  been measured, being  mostly  outside the   overlapping zone of the  VLT and HST fields. \\

In the field around the  six-image lensed quasar {\bf \btreize\,} we have measured  the photometric redshifts of  galaxies appearing in the overlapping  VLT and HST fields (see Fig.\ref{histo}). There are very few galaxies for which the SED is fitted with a $\chi^2 \sim$1 (16 galaxies among the 34 detected in the four bands). Half of them have a redshift between 1.1 and 1.5 and  might correspond to the  galaxy overdensity detected towards  \btreize\, because of their location in the field. Therefore, we will consider in the following, that the redshift of the galaxy cluster towards \btreize\, is $z=$1.3.\\

In the field around {\bf \clover\,} we have computed the photometric redshifts for galaxies located in the area covered by both the VLT and HST datasets. Two peaks are observed in redshift: a first one at $z=$0.8$\pm$0.3, and a second one,  at $z=$1.75$\pm$0.2. As the two overdensities are mostly superimposed on the common HST and VLT frame, it is hard to identify their corresponding redshift. Yet, in the original paper reporting on the discovery of the  South-East overdensity discovery  (Kneib et al. 1998b), the  photometric redshift derived (from three filters only)  was $z$= 0.9$\pm$0.1. We will assume in the following that the redshift of the South-East overdensity is $z$=0.8$\pm$0.3. The second galaxy overdensity could correspond either to the absorption system at $z$=1.661 or to that at  $z$=1.87 (Turnshek et al. 1998; Magain et al. 1998.)  \\

Finally, towards {\bf  \hevingt\,}  the distribution of photometric redshifts (using the HST and VLT images) shows a peak in  redshift at $z=$0.7$\pm$0.1 with a high confidence level. This suggests the presence of a galaxy cluster or galaxy group towards the doubly imaged quasar. On the other  hand, previous studies by Wisotzki et al. (1998) and Lopez et al. (1998) argued that the lensing galaxy maximal redshift is $z=$0.5, according to its V-R color. Due to the uncertainty on the photometric redshift determination and on the redshift limit given by Wisotzki et al., it is still  uncertain whether the lensing galaxy is part  of the galaxy cluster/group discovered in this study. \\

\subsection{Color versus magnitude diagrams}
We have analysed the  R-Ks color versus Ks magnitude  diagrams of the galaxies detected in the overlapping FORS1 and ISAAC fields. Unfortunately, no elliptical red sequence appears in any of the field.  The presence of such a sequence would argue in favor of the presence of a massive galaxy cluster towards the quasar. There are  possibly two  reasons for a non-detection: 1) there are too few galaxies in the  common $R$- and $Ks$-band  catalogs (between 50 and 200). Hence, the elliptical red sequence of galaxy clusters does not show up in a conspicuous manner. 2) the galaxy clusters or groups are not massive enough to have an  elliptical red sequence.  

\section{Weak-shear analysis and mass reconstruction}\label{weakmass}

\begin{table*}
\renewcommand{\arraystretch}{1.0}
\centering
\begin{center}
\caption{Results of the weak-shear  analysis and of the mass reconstruction. Column 1: quasar name. Column 2: number of galaxies per arcmin$^2$ corresponding to the galaxy overdensity. Column 3: position of the galaxy overdensity center relative to the quasar images. Column 4: photometric redshift of the galaxy overdensity and  corresponding error bar (``$^*$'': value inferred from previous paper, see text for references). Column 5: magnitude range for the background galaxies used for the weak-shear analysis . Column 6: number of galaxies used for the weak-shear analysis and the mass reconstruction. Column 7: ICF width of the mass reconstruction in arcsecond. Column 8: upper limit of the mass integrated from the mass reconstruction  in a radius of 500~kpc and centered at the galaxy group position. Column 9: galaxy cluster  velocity dispersion corresponding to the mass given in Column 8. Column 10: shear orientation in degree. }
\begin{tabular}{|c|l|l|l|l|l|l|l|l|l |}
\hline
Quasar   &Nb gal &  Relative   &z$_{phot}$ & Sources&  Nb gal & ICF  &   Mass        & $\sigma_v$ & PA$_\gamma$ \\  
                & /(\arcmin)$^2$ &   position &$\pm \Delta$z&  Mag. range  & for WS & ('')           &  (10$^{14}$M$_\odot$h$^{-1}$)& (km~s$^{-1}$h$^{-1}$) &\\
      
\hline
CTQ~414 &28&  12\arcsec SE & 0.5$\pm$0.1 &25$<$R$\leq$28  &578 &180   &$<$1.6& $<$720 & 24$^o$ \\
\hline
HE~0230-2130  &21.1 & 38\arcsec W& 1.0$^\star $&25$<$R$\leq$28 &633 &120  &4.0 & 1130 &45$^o$ \\
\hline 
\lbqs  &\_ &\_&1$^\star $&25$<$R$\leq$28&351 &100&  $<$2.4& $<$880 & -33$^o$\\
\hline
\bdix  & 10 &120\arcsec SE&0.6$^\star $ &23$<$R$\leq$28&416 &120  &$<$2.3  & $<$860  & 33$^o$ \\  
\hline
HE~1104-1805 &9 &40\arcsec N & 2.0$\pm$0.1 &24$<$R$\leq$28&\_ &\_  & \_&\_ & 90$^o$\\ 
             & 18 & 78\arcsec SW & \_& \_&  \_ &\_  & \_ & \_&\_\\
\hline
B~1359+154   &4 &7\arcsec NE& 1.3$\pm$0.2 &23$<$R$\leq$28& 749&300  & $<$10.7 & $<$1800 & -12$^o$\\
\hline 
\clover &43 & 14\arcsec SE &0.8$\pm$0.3 & 25$<$R$\leq$28&453 & 100  &$<$1.8 & 760 & 45$^o$\\
        & 32 &  36\arcsec SW & 1.7$\pm$0.2 & \_&\_ &\_ &\_ &\_ &\_ \\
\hline
HE~2149-2745 & 22 & 7\arcsec SE& 0.7$\pm$0.1 & 25$<$R$\leq$28&81&60 &3.2&1010 &45$^o$\\
\hline
\end{tabular}
\label{result}
\end{center}
\end{table*}

For each field, we have searched  for the weak-shear signature that the galaxy groups or clusters should imprint on background galaxies. According to the gravitational lensing theory, the presence of a  galaxy cluster or a galaxy group in the field generates  distortions and amplifications of the  background  galaxies. In practice, we measure the ``statistical'' distortions:  the  mean  ellipticity and mean position angle  of the faintest galaxies in  cells of a given size. \\
\subsection{PSF correction}
We have performed the weak-shear analysis on the FORS1/$R$-band  images of the systems.
First of all, we have to correct the galaxy ellipticities for the  instrumental distortions across the field. Each night,  the   PSF variations  across the field are measured from a sample of stars, using the {\it  Im2shape} software  (Bridle et al. 2002, Kneib et al. 2003). 
The ellipticity of the PSF varies typically from $\epsilon$=(a-b)/(a+b)=0.002 to 0.05 across the FORS1 field. 
 Then, the background galaxies  are deconvolved in an analytical way, modeling them by   2D-functions and   using the  closest  PSF. 
\\
\subsection{Distorsion measurement}
Once the galaxy images have been deconvolved and  using only  galaxies with ellipticities known with an accuracy better  than 20\%, we can compute  mean galaxy ellipticities and mean galaxy orientations across the field. The field is divided into   a grid of cells. In order to optimize the coherence of the shear-map pattern, we attribute to any galaxy within the cell an ellipticity and an orientation derived through averaging the  ellipticities and the orientations of all galaxies
 within a radius $r$ slightly larger than half the cell size. Then, within each  cell, the attributed ellipticities and attributed  orientations are averaged for all the galaxies. These final ``mean ellipticity'' and  ``mean orientation'' are allocated to the cell center.  
 In a field with a massive cluster (Cl~0024+1654, Hudelot et al. 2004, private communication)  we have tested that,  for the total number of galaxies used to perform the weak-shear analysis (between 1000 and 2500 galaxies per field), a combination of a cell $\sim$40\arcsec$\times$40\arcsec, and   a  radius $r$  of 0.01$^{\rm o}$ to 0.02$^{\rm o}$, allows to extract the statistical information in a satisfactory way. \\

None of the shear map of these  systems reveals a  signal strong enough to be spotted at first glance. However the shear map can be  used to derive the lensing potential mass distribution as well as an upper limit of its mass. This is detailed in the following subsection.



\subsection{Mass reconstruction}

\begin{figure*}
\begin{center}
\caption{\label{ws2} Mass reconstruction superimposed on the FORS1 images. The contours outline  regions of same signal-to-noise (dash-lines: S/N$<$1, solid-lines: S/N$\ge$1). From the top left hand corner to the bottom right hand corner: (a) \ctq, (b) \hezero, (c) \lbqs, (d) \btreize\, and (e) \clover.}
\end{center}
\end{figure*}

 The galaxy mean ellipticities  and mean orientations  are given as input to the  {\it LensEnt2} software (Bridle et al. 1998, Marshall et al. 2002), a code which allows to built the  mass distribution of a lens using the shear-map.  The redshift of the lens plane  assumed for this reconstruction is that provided in Table \ref{result} (column 4). The source plane redshift is derived from the photometric redshift of the FDF galaxies (E. Labbe, private communication, see Heidt et al. 2003 for the photometry): in a given magnitude range (displayed in Table \ref{result}), we have averaged the FDF galaxy photometric redshifts and assigned this mean redshift  to the galaxy sample. For all magnitude ranges considered here,  the mean galaxy  redshift is  $z \sim$ 1.4.  The reconstruction is performed in a frame 4 times larger than the  frame where the weak-shear was analysed since the shear pattern eventually observed  in the FORS1 field could be due to the presence of a mass potential located outside the field of view.    {\it LensEnt2} uses a maximum entropy method to reconstruct the mass potential. The program simulates  a synthetic  population of unlensed background galaxies at a  same redshift and in a field of the same size as the observed one  (Bartelmann et al. 1996). Then {\it LensEnt2} adds in the field a mass clump, with a compactness characterized by its  Intrinsic Correlation Function (ICF, Marshall et al. 2002), at the lens redshift, and analyses the similarity between this simulated  shear pattern and the observed shear pattern  in each  field.\\ 
   The ICF of the mass clump is assumed to be Gaussian. For each system, we have performed a set of mass reconstructions using different ICF widths. The "evidence" (see Marshal et al. 2002) determined for each mass reconstruction  provides a good discriminator of the ICF width, weighting the reality of the obtained mass reconstruction. The width of the ICF finally selected is given in Table \ref{result}. Typically, the  mass bump which reveals a galaxy cluster in a reconstruction map corresponds to an ICF width between 100\arcsec\, and 300\arcsec.\\
      Since the  noise properties  of maximum-entropy inversion  methods are
not easy to interpret, we have performed, for each system,  80 reconstructions starting with  modified
catalogs of  galaxies (positions are retained  but ellipticity
directions  are assigned  at  random). For each of the 80 
mocked catalogs, mass reconstruction is performed, and a set of mass bumps is obtained. Integrating over the 80 catalogs, the mass bump distribution shows a number of peaks. Their median gives the noise level of this procedure.
 Results of the mass reconstructions  towards the  systems  are displayed in Fig. \ref{ws2}. The contours outline  regions of same signal to noise ratio. The signal to noise of the mass reconstruction is low  at the location of the galaxy overdensity (S/N$<$3 in all frames), and there are regions in the field with higher mass signal than at the position of the galaxy overdensity.  The low signal at the overdensity location could have different origins: 1) there is actually no mass clump at this position 2) the mass inferred in the fields are too low to be detected by the reconstruction, and therefore the masses measured here are upper limit 3) the largest structures in the field (real or artifacts due to the low number of background galaxies in some cases) are adding noise, and the signal level at the overdensity position is biased. Therefore, we investigate the limit of the {\it LensEnt2} software performing the following test: for each system, we have lensed the background galaxies with a Pseudo Isothermal Elliptical Sphere (Kassiola \& Kovner 1993) of small ellipticity ($\epsilon$=(a$^2$-b$^2$)/(a$^2$+b$^2$)=0.05) and centered on the quasar image A. We have looked for the smallest mass potential that the software can reconstruct. This value varies in each field as the number of background galaxies is different. The lower mass values (in a radius of 500kpc) are displayed in Table \ref{mass} (column3), as well as the mass integrated in a same radius, but centered on the overdensity in the original mass map.

\begin{table}
\renewcommand{\arraystretch}{1.0}
\centering
\begin{center}
\caption{\label{mass}Galaxy group masses. Column 1: quasar name. Column 2: mass integrated in the original mass map, in a radius of 500~kpc centered on the galaxy overdensity. Column 3: lower mass value reconstructible by {\it LensEnt2}.}
\begin{tabular}{|c|l|l|}
\hline
 & M($<$500~kpc) & M$_{lensed}$($<$500~kpc) \\
 &10$^{14}$M$_\odot$h$^{-1}$ &10$^{14}$M$_\odot$h$^{-1}$\\
\hline 
\ctq &  1.6 & 1.7 \\
\hezero & 4.0 & 3.6 \\
\lbqs & 2.4 & 5.3 \\
\bdix & 2.3 & 2.4 \\
\btreize & 10.7 & 12.7 \\
\clover & 1.8 & 2.0 \\
\hevingt & 3.2 & 2.6 \\
\hline
\end{tabular}
\end{center}
\end{table}


\subsection{Results}

\begin{figure}
\begin{center}
\includegraphics[width=8cm]{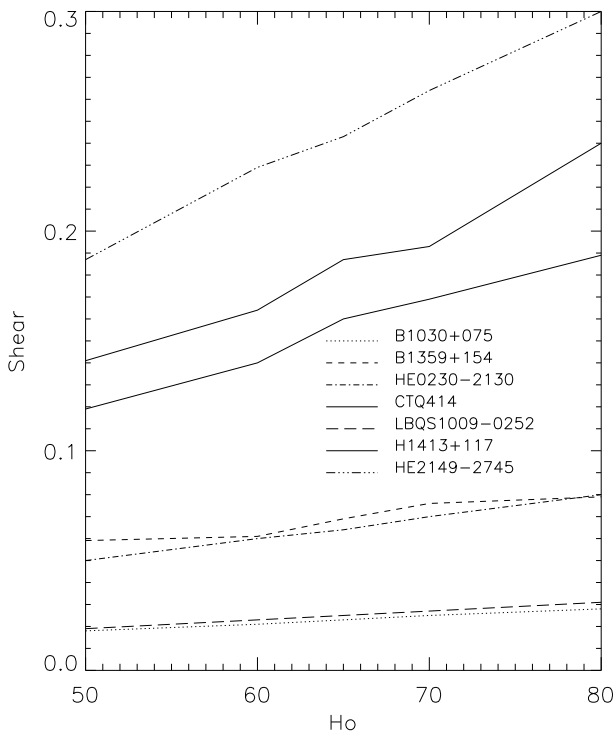}
\caption{\label{shearvar} Shear strength induced by the galaxy groups towards the different systems as a function of H$_0$.}
\end{center}
\end{figure}
The sizes of the mass clumps computed by the program, as well as the masses inferred, correspond to small galaxy clusters, or even galaxy groups. 

 In Table \ref{result}, the mass displayed are   integrated over  a radius  of 500~kpc (typical size for a galaxy cluster) and centered on the galaxy overdensity. The  velocity dispersion is derived from the mass value,  assuming that the mass distribution follows an  Isothermal Sphere potential  with a core radius of 30~kpc.  

We have chosen to rebuild the mass towards \lbqs\, even though no evidence for the presence of any galaxy cluster towards the quasar has been found. We choose arbitrarily $z$=1 for the  lensing potential redshift and a galaxy magnitude range of R=[23,28[ for the measurement of distortions. The lack of mass concentration confirms that we do not detect any extra gravitational potential in the field around \lbqs.    \\

From the results of the mass limit test displayed in Table \ref{mass}, we see that the only systems  for which the mass inferred  M  have a physical relevancy are  \hevingt\, and \hezero, as they appear to be larger  than the detection limit M$_{lensed}$. In the other cases the  masses have to be taken as upper mass limits. In the case of \lbqs\, the mass measured  is well below  the detection limit, and this suggest  that the overdensity detected here is not related to a mass potential.\\

Finally, the  {\it LensTool} software  provides a map of the shear induced by a mass distribution according to the distance to the potential center. Therefore, we can easily extract the shear strengths that the galaxy groups induce at the quasar positions.  These values are  displayed in Fig. \ref{shearvar} versus the value of  the Hubble constant.

The shears computed here are derived from the  masses of  the galaxy groups. They are, as a consequence, shear  upper  limit values in the same case of \ctq, \bdix, \btreize\, and \clover. We notice that they are in general lower than those  predicted in former studies (displayed in Table \ref{castles}). Moreover the orientations are also different from  the predicted ones. These discrepancies between our results and  previous  ones might come from the fact that the predicted shear values and  orientations had been formerly deduced by modeling the lensing galaxy as a sphere, that is to say  without taking into account its orientation and ellipticity, while these last parameters dominate the shear strength and orientation in most cases.

\section{Conclusions}\label{discussion}
We have studied the surroundings of the quasar images in eight gravitationally lensed systems in order to search for the presence of galaxy clusters or galaxy groups which could contribute to the ``external shear'' requested from  previous studies. We have used  three different and independent methods: 
the search of galaxy number overdensities, the search of a peak in the redshift distribution of galaxies -using photometric redshifts-, and finally a weak-shear analysis combined with mass reconstruction of the background galaxies.\\
There is convincing evidence for the presence of a galaxy group in the line of sight towards the following 5  lensed quasars: \ctq, \hezero, \btreize, \clover\, and \hevingt. Towards \lbqs\, and \bdix\, we have not been able to compute the photometric redshift of the overdensity, and, therefore, we cannot  conclude about their 3D reality. The galaxy overdensity discovered in the field around  \heonze\, seems to be associated with the quasar rather than with the lensing potential.\\

The mass reconstruction leads to small mass clump sizes, related to  galaxy groups. Moreover, the galaxy group masses and the shear they induce at the quasar position have  to be regarded as  upper limits, as we are  using the reconstruction mass method at the edge of its applicability. 

The next step of this work is to incorporate the information derived from the wide-field  analyses in the  modeling of the lensing potential towards these eight lensed quasars. This will ultimately lead to an improved value for the Hubble constant, as derived  from  available time-delay measurements between the quasar image light-curves.

 \begin{acknowledgements}
 The HST data used  in this paper were  obtained by the
``CfA  Arizona Space Telescope  LEns Survey''  (CASTLES) collaboration
(PI:  E.  Falco).   C.F.  acknowledges support  from  an ESO
studentship in  Santiago and a grant from the ``Soci\'et\'e de Secours des Amis des Sciences''. F.C. is supported by the European Comission through Marie Curie grant MCFI-2001-0242. The ECOS/CONICYT grant CU00U05 is also gratefully acknowledged. J-P.K. acknowledges support from Caltech and CNRS. F.C. is partially funded by the ``P\^ole d'Attraction Interuniversitraire'' P4/05 (STSC, Belgiun).
\end{acknowledgements}

\end{document}